# Multi-Label Chest X-Ray Classification via Deep Learning

## Aravind Sasidharan Pillai

University of Illinois Urbana-Champaign, Champaign, IL, USA
Email: pillai.aravinds@gmail.com

## Abstract

In this era of pandemic, the future of healthcare industry has never been more exciting. Artificial intelligence and machine learning (AI & ML) present opportunities to develop solutions that cater for very specific needs within the industry. Deep learning in healthcare had become incredibly powerful for supporting clinics and in transforming patient care in general. Deep learning is increasingly being applied for the detection of clinically important features in the images beyond what can be perceived by the naked human eye. Chest X-ray images are one of the most common clinical method for diagnosing a number of diseases such as pneumonia, lung cancer and many other abnormalities like lesions and fractures. Proper diagnosis of a disease from X-ray images is often challenging task for even expert radiologists and there is a growing need for computerized support systems due to the large amount of information encoded in X-Ray images. The goal of this paper is to develop a lightweight solution to detect 14 different chest conditions from an X ray image. Given an X-ray image as input, our classifier outputs a label vector indicating which of 14 disease classes does the image fall into. Along with the image features, we are also going to use non-image features available in the data such as X-ray view type, age, gender etc. The original study conducted Stanford ML Group is our base line. Original study focuses on predicting 5 diseases. Our aim is to improve upon previous work, expand prediction to 14 diseases and provide insight for future chest radiography research.

## 1. Introduction

Radiology is a branch of medicine that can be divided into diagnostic radiology and interventional radiology [1]. Diagnostic radiology involves examining the medical images to diagnose diseases and abnormalities. Chest X-ray radiography is the most common imaging examination that demands correct and immediate interpretation to avoid life-threatening diseases. The challenge arises when these images have to be interpreted by radiologists who are limited by speed, experience and the cost involved to get a certified radiologist. Therefore, health care industry turned towards deep learning algorithms to automate and generate accurate radiology reporting.

Deep learning in health care is bursting with possibility and remarkable innovation by providing the ability to analyze vast quantities of data at exceptional speed without compromising on accuracy [2]. It enables the creation of algorithms that can learn and make predictions. In contrast to rules-based algorithms, machine learning takes advantage of increased exposure to large data sets and possesses the ability to improve performance with such exposures and learn with experience.

Deep learning is a subset of artificial neural networks that are statistical and mathematical methods inspired by the way biological nervous system processes information with a large number of highly connected neurons, nodes or cells. Neural networks are structured as one input layer, one or more hidden layers and an output layer. Every hidden layer consists of a set of neurons that are fully connected to all neurons in the previous layer. The strength of such a connection is determined by weights of variable or features that associate inputs with outputs. For a neural network model to perform efficiently and accurately, these weights must be set to suitable values, which are estimated through training.

From deep learning perspective, radiology images need to be pre-processed differently due to variations in processors and memory restrictions. X-ray images in general are 2D images of a 3D human body. DL algorithms especially convolutional neural networks (CNN) have proved more successful in training the models with 2D images than 3D images that adds an additional dimensionality to the problem. Convolution is a mathematical operation that employs a type of filtering to determine the most useful features from a dataset, thereby having applications in finding patterns in signals or filtering signals.

## 1.1. Related Works

In recent years, large sets of radiology images have been made public. The availability of such datasets helped crowd-source the development and evaluation of deep learning models [3] [4] [5] [6]. In 2017, NIH Clinical Center released over 100,000 chest x-ray images, which comprises 108,948 frontal-view X-ray images of 32,717 unique to the scientific community. There were several studies based on this data. Wang *et al.* [7] through their paper "ChestX-ray8: Hospital-scale Chest X-ray Database and Benchmarks on weakly-Supervised Classification and Localization of Common Thorax Diseases" demonstrated that the commonly occurring thoracic diseases can be detected and even spatially located via a unified weakly supervised multi-label image classification and disease localization framework.

In 2019, MIT published MIMIC-Chest X-Ray Database (MIMIC-CXR) collection of more than 350,000 chest x-ray images associated with 227,943 studies. These images consist of both frontal and lateral views and were compiled from Beth Israel Deaconess Medical Center in Boston during years 2011 to 2016. Images are provided with 14 labels derived from free-text radiology reports using NLP tools [8].

CheXpert is another such dataset consisting of 224,316 chest radiographs of 65,240 patients who underwent a radiographic examination from Stanford University Medical Center between October 2002 and July 2017, in both inpatient and outpatient centers. Based on associated radiology reports, these X-rays images were labeled as positive, negative, or uncertain for the presence of 14 common chest radiographic observations. CheXpert data has attracted strong attention in building pre-trained learning models to address the challenges in X-Ray image processing, classification and segmentation. CheXNet is one prominent project by Stanford ML Group which is considered to have state of the art performance in classifying diseases even better than expert radiologists.

## 1.2. Baseline

In this paper, we take inspiration from state-of-the-art CheXNet model to train similar models based on CNN that perform multi-class, multi-label classification with transfer learning from models trained on Imagenet data. Irvin *et al.* [1], in their paper developed models to predict five lungs' conditions such as Atelectasis, Cardiomegaly Consolidation, Edema and Pleural Effusion. Their study achieved AUC ranging from 0.85 to 0.93. Our goal is to expand the predictions from 5 diseases to 14 diseases and achieve comparable AUCs, and also attempt to compare various pre-trained CNN models and their performance.

## 2. Approach and Methods

Our approach primarily focuses on developing CNN classifier to predict the labels. During research, it is a good practice to evaluate models on different datasets since deep learning algorithms are often validated on historical data, yielding guaranteed performance but are unable to attain same levels of accuracy when operating outside the training data range. This is called as overfitting. We employ a method to compare the performance of different models based on success metrics of the model on test data set of images and determine the best performing model under different parameter setting.

### 2.1. Convolutional Neural Network

CNN is a deep Learning algorithm which can take an input image, assign importance as weights to various features in the image and be able to differentiate one from the other. The main advantage of Convolutional Neural network is that it has the capability to capture the temporal and spatial dependencies in an image by applying relevant filters [9]. CNN architecture is composed of a convolutional layer, a pooling layer and a fully connected layer. The function of a convolutional layer is to extract features from images. Each convolutional layer can have multiple convolution kernels and the convolutional layer is calculated as follows:

$$x_j^l = f\left(\sum_{i \in M_j} x_j^{l-1} * k_{ij}^l + b_j^l\right),$$

where $x_j^{l-1}$ is the characteristic map of the output of previous layer, $x_j^l$ is the output of the $i$th channel of the $j$th convolutional layer and $f$(.) is called the activation function. $M_j$ is the subset of input feature maps, $k_{ij}^l$ is a convolutional kernel and $b_j^l$ is its corresponding weight. A pooling layer sits between two convolutional layers to reduce the feature map dimensions and still maintain the scale of the features to an extent. Mean pooling and Max pooling are two main pooling methods used. A fully connected layer integrates multiple image maps after they are passed through multiple convolutional and pooling layers to obtain high-level semantics used to determine the class labels.

All the experiments in our work involve millions of parameters for multiple features and multiple layers of neural networks. This requires us to use a faster processing machine with graphic processing units (GPU) and cloud architecture that have pre-configured drivers and come with popular Python packages to build neural networks.

## 2.2. Experimental Setup

We used AWS Deep Learning AMI (Ubuntu 18.04) for training and prediction. These AMIs are pre-configured with PyTorch 1.7.1 and Python3.7 (CUDA 11.1 and Intel MKL) along with other basic data science software like pandas, numpy, scipy etc. We used g4dn.2xlarge machine with 1 Tesla T4 GPU.

## 2.3. Dataset

We used a version of Chexpert data with down sampled resolution to train the model. The original data is 439 GB in size whereas the down sampled version is just 11 GB. We chose the down sampled version to save training time and computing power. Sample images are shown in Figure 1.

Each individual X-ray radiograph is represented by the image and a feature vector containing patient id, sex, age, study number, X-ray view-type, and a vector of fourteen expert-labeled observations. We clean the data to omit the sex and age features from the equation because we evaluate solely on the image.

### Train Data Distribution

This subset of dataset is split into train and validation sets for model evaluation with a split proportion of 80:20. Figure 2 and Figure 3 denote the distribution of train data and train labels respectively. An unseen test data set is used to evaluate predictions of classes by the trained models.

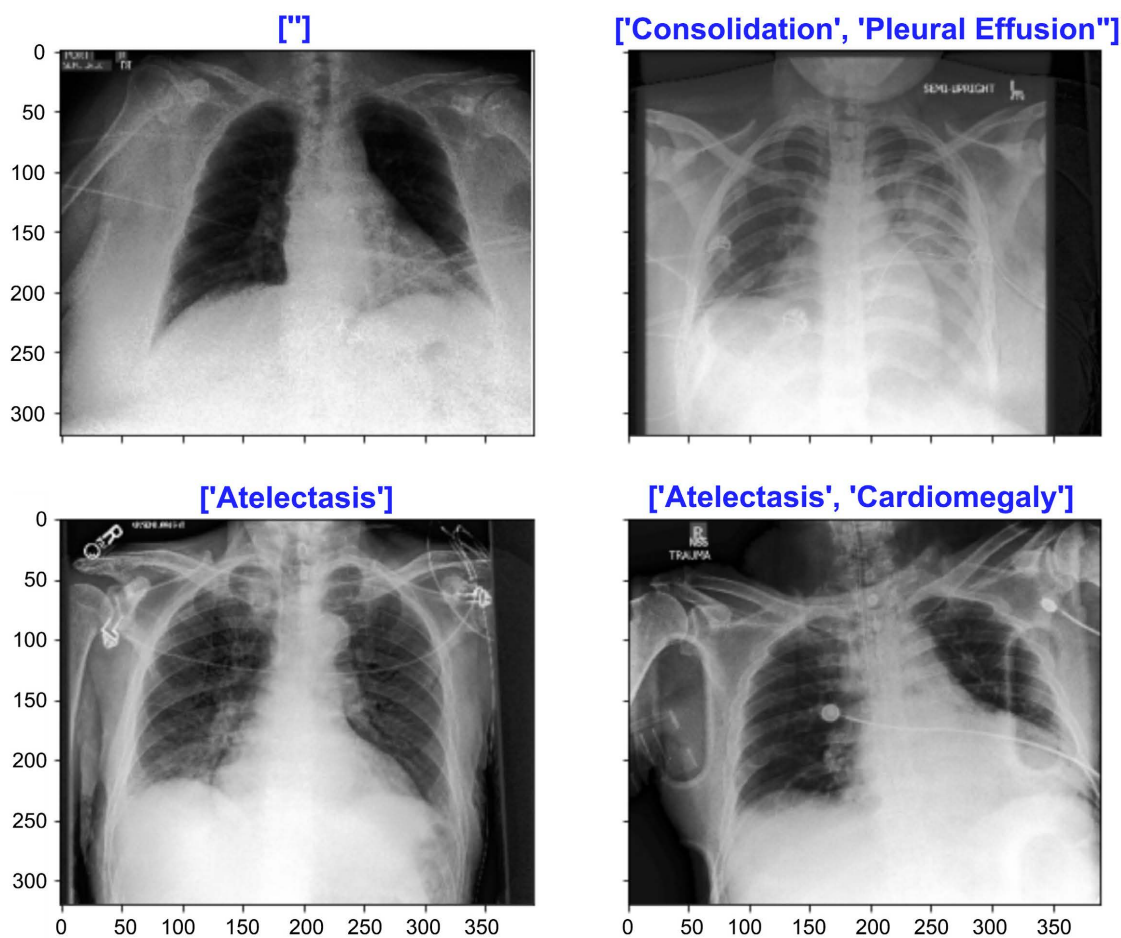

**Figure 1.** Sample images.

| | Label | minusOneVal | oneVal | zeroVal | nanVal |
|---|---|---|---|---|---|
| **0** | No Finding | 0 | 22381 | 0 | 201033 |
| **1** | Enlarged Cardiomediastinum | 12403 | 10798 | 21638 | 178575 |
| **2** | Cardiomegaly | 8087 | 27000 | 11116 | 177211 |
| **3** | Lung Opacity | 5598 | 105581 | 6599 | 105636 |
| **4** | Lung Lesion | 1488 | 9186 | 1270 | 211470 |
| **5** | Edema | 12984 | 52246 | 20726 | 137458 |
| **6** | Consolidation | 27742 | 14783 | 28097 | 152792 |
| **7** | Pneumonia | 18770 | 6039 | 2799 | 195806 |
| **8** | Atelectasis | 33739 | 33376 | 1328 | 154971 |
| **9** | Pneumothorax | 3145 | 19448 | 56341 | 144480 |
| **10** | Pleural Effusion | 11628 | 86187 | 35396 | 90203 |
| **11** | Pleural Other | 2653 | 3523 | 316 | 216922 |
| **12** | Fracture | 642 | 9040 | 2512 | 211220 |
| **13** | Support Devices | 1079 | 116001 | 6137 | 100197 |

**Figure 2.** Distribution of train data.

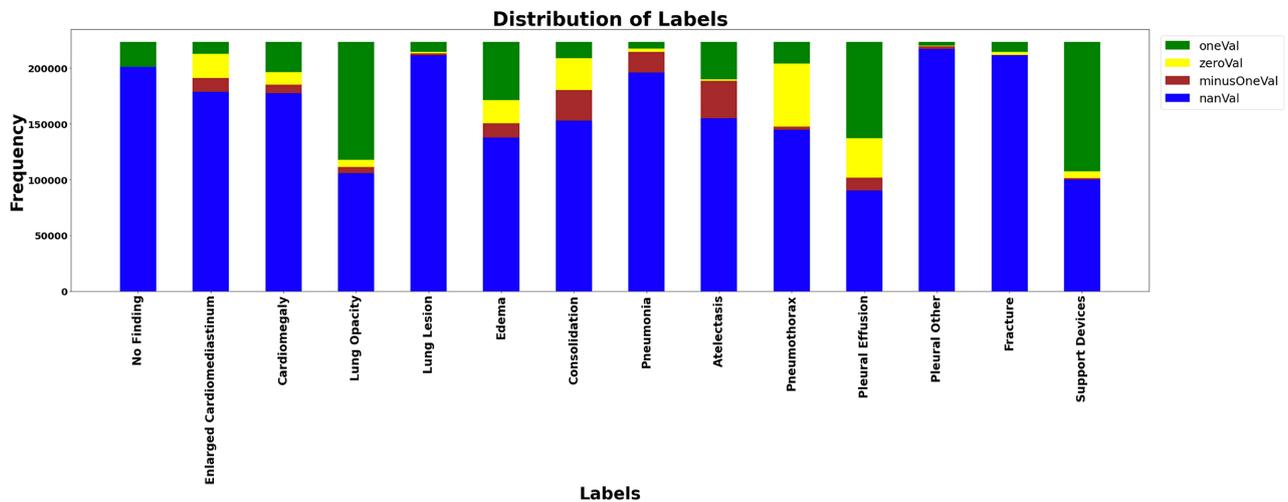

**Figure 3.** Distribution of train labels.

### 2.4. Data Uncertainty

Chest X-ray dataset contains images that are hard to classify for a certain disease to be present. So, radiologists often leave an uncertain label (value = −1) on such images. We followed approach similar to Irvin *et al.* [4] to handle uncertainties. Labels were split into u-zero and u-one categories based on previous studies. Category u-one means uncertain labels will be treated as positive and u-zero means negative. We considered labels Atelectasis, Edema as u-ones and rest of the labels as u-zero.

### 2.5. Data Pre-Processing

Dataset used for training and test are pre-processed through augmentation methods. This was done to increase the size and quality of the dataset. This process helps in solving problems related to overfitting and enhances the model's generalization ability during training and prediction of class labels.

### 2.6. Modeling

Based on our research, CNN architecture performs better on multi-class, multi-label classification of image dataset due to the reduction in number of parameters involved, without losing features that are critical for getting a good prediction. So, we investigated multiple models based on CNN architecture that will be discussed in detail further. Each of this model is run on train and test data with a batch size of 96 for up to 40 epochs, binary cross entropy as loss function, Adam optimizer with an initial learning rate of 0.001 which is multiplied by 10 each time the validation loss plateau after an epoch.

#### 2.6.1. Custom Net (Simple Baseline Model)

We started with a simple custom-made CNN models. Trained the model from scratch where input is passed through 4 convolutional layers by random initialization and fine tuning the weights of all the layers. Each convolutional layer has a max pool layer and ReLU activations. Pooling layers used to reduce spatial volume. Figure 4 demonstrates list of parameters for each layer. Output of these layers is then passed to a fully connected sigmoid function to classify a collection of chest X-ray images. Sigmoid function helps to convert raw output values from classifier to corresponding probability values. We considered probability greater than 0.50 as positive detection.

### 2.6.2. DenseNet121

The core idea of DenseNet is to ensure maximum information flow between layers in the network by connecting all layers directly with each other. It has a stack of dense blocks followed by transition layers. Dense blocks contain different units such as convolutions, batch normalization and ReLU activations. Each dense block generates a fixed number of feature vectors which is called the Growth rate *i.e.*, the amount of information that layers can transmit. We trained DenseNet121 with initial weights from a pre-trained network on ImageNet data [10].

### 2.6.3. ResNet-50

ResNet-50 is a convolutional neural network that is 50 layers deep. We initialize the network with pre-trained weights, where knowledge is transferred from ImageNet data. Pre-trained network initial weights are frozen for first 6 layers used for feature extraction and only the weights of the last layer are adapted to re-train one or more layers with samples from the X ray dataset [11].

### 2.6.4. Inception_V3

This network has 48 layers depth that can make several improvements including label smoothing, factorized convolutions and uses an auxiliary classifier to propagate label information lower down the network. We initialize the network with

```
CustomNet
+-----------------------+------------+
|        Modules        | Parameters |
+-----------------------+------------+
|   ConvLayer1.0.weight |    216     |
|    ConvLayer1.0.bias  |     8      |
|   ConvLayer1.1.weight |   1152     |
|    ConvLayer1.1.bias  |     16     |
|   ConvLayer2.0.weight |   12800    |
|    ConvLayer2.0.bias  |     32     |
|   ConvLayer2.1.weight |   9216     |
|    ConvLayer2.1.bias  |     32     |
|   ConvLayer3.0.weight |   18432    |
|    ConvLayer3.0.bias  |     64     |
|   ConvLayer3.1.weight |   102400   |
|    ConvLayer3.1.bias  |     64     |
|   ConvLayer4.0.weight |   204800   |
|    ConvLayer4.0.bias  |    128     |
|   ConvLayer4.1.weight |   147456   |
|    ConvLayer4.1.bias  |    128     |
|       Lin1.0.weight   |   7168     |
|        Lin1.0.bias    |     14     |
+-----------------------+------------+
Total Trainable Params: 504126
```

**Figure 4.** CustomNet training parameters.

pre-trained weights, where knowledge is transferred from ImageNet data and freeze these weights for first 8 layers [12].

### 2.6.5. Vgg16

The VGG network is a neural network that has already been pretrained on over a million images from the ImageNet database. The network has 41 layers. There are 16 layers with learnable weights, 13 convolutional layers, and 3 fully connected layers. One of the major disadvantages of the VGG16 Neural Network is the huge number of trainable parameters. It has more than 134 million trainable parameters. We did freeze the first 6 layers so as to limit the trainable parameters to 57 k [11] [13].

Figure 5 demonstrates trainable parameters for each of these models.

### 2.7. Success Metrics

**Accuracy**: It is the ratio of number of correct predictions to the total number of input samples.

**Area Under Curve**: AUC of a classifier is equal to the probability that the classifier will rank a randomly chosen positive example higher than a randomly chosen negative example [14].

**F1 Score**: F1 Score is the Harmonic Mean between precision and recall.

**Precision**: The precision is the ratio the number of true positives to the sum of true positives and false positives. The precision denotes the ability of the classifier not to label as positive a sample that is negative.

**Recall:** The recall is the ratio the number of true positives to the sum of true positives and false negatives. The recall denotes the ability of the classifier to find all the positive samples.

### 2.8. Model Training

We did a random 20% split of train data as validation data and trained model for a maximum of 40 epochs. ROC was used to determine early stopping criteria. Most of the models achieved best performance with 20 to 25 epochs. Figure 6 demonstrates variations in training metrics per epoch.

DenseNet model achieved highest training AUROC 78 and highest training accuracy 87% as indicated in Figure 7.

## Training Parameters

|   | Model Name | Total Parameters | Trainable Parameters |
|---|---|---|---|
| 0 | CustomNet | 504126 | 504126 |
| 1 | DenseNet121 | 6968206 | 6968206 |
| 2 | ResNet50 | 23772110 | 264078 |
| 3 | Inception | 21814254 | 28686 |
| 4 | Vgg16 | 134317902 | 57358 |

**Figure 5.** Training parameters summary.

## Training Metrics Comparison

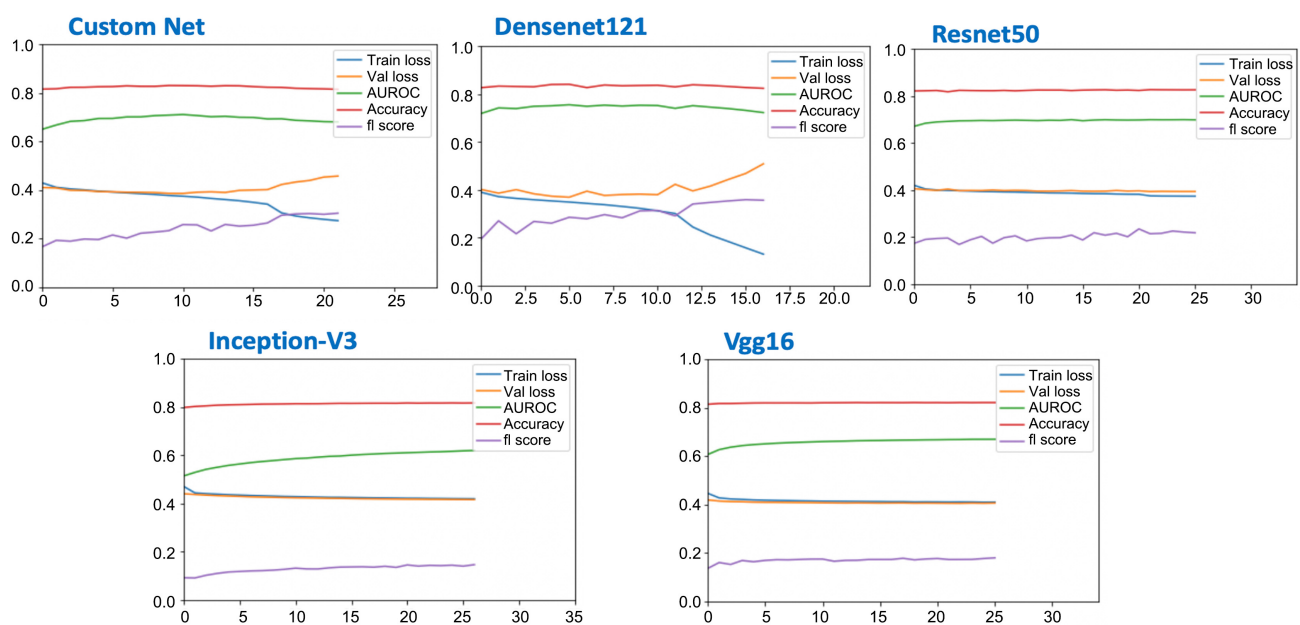

**Figure 6.** Training metrics per epoch.

## 2.9. Training and Validation Metrics

| | Model Name | AUROC | Accuracy | f1 Score |
|---|---|---|---|---|
| **0** | CustomNet | 0.740989 | 0.862637 | 0.259044 |
| **1** | DenseNet121 | 0.779881 | 0.866486 | 0.271429 |
| **2** | ResNet50 | 0.716902 | 0.854088 | 0.219912 |
| **3** | Inception | 0.650021 | 0.846562 | 0.151924 |
| **4** | Vgg16 | 0.671470 | 0.848543 | 0.164998 |

**Figure 7.** Training metrics summary.

## 3. Results

To evaluate the performance of models, we used unseen test data to predict the multi classifications labels. The validation set contains 200 studies from 200 patients randomly sampled from the full dataset with no patient overlap with the train set. Test data consisted of 234 images. Distribution of test data and labels is shown in **Figure 8**.

## 3.1. Test Metrics

Model performance validated against success metrics. Details available are in **Figure 9**.

Based on overall test metrics, DenseNet121 achieved the best performance. This model achieved ROC score of 0.78 and accuracy of 87%. Other models ROC values ranged from 0.69 to 0.75 and accuracy from 83% to 86%.

## Test Dataset

|    | Label | minusOneVal | oneVal | zeroVal | nanVal |
|----|-------|-------------|--------|---------|--------|
| 0  | No Finding | 0 | 38 | 196 | 0 |
| 1  | Enlarged Cardiomediastinum | 0 | 109 | 125 | 0 |
| 2  | Cardiomegaly | 0 | 68 | 166 | 0 |
| 3  | Lung Opacity | 0 | 126 | 108 | 0 |
| 4  | Lung Lesion | 0 | 1 | 233 | 0 |
| 5  | Edema | 0 | 45 | 189 | 0 |
| 6  | Consolidation | 0 | 33 | 201 | 0 |
| 7  | Pneumonia | 0 | 8 | 226 | 0 |
| 8  | Atelectasis | 0 | 80 | 154 | 0 |
| 9  | Pneumothorax | 0 | 8 | 226 | 0 |
| 10 | Pleural Effusion | 0 | 67 | 167 | 0 |
| 11 | Pleural Other | 0 | 1 | 233 | 0 |
| 12 | Fracture | 0 | 0 | 234 | 0 |
| 13 | Support Devices | 0 | 107 | 127 | 0 |

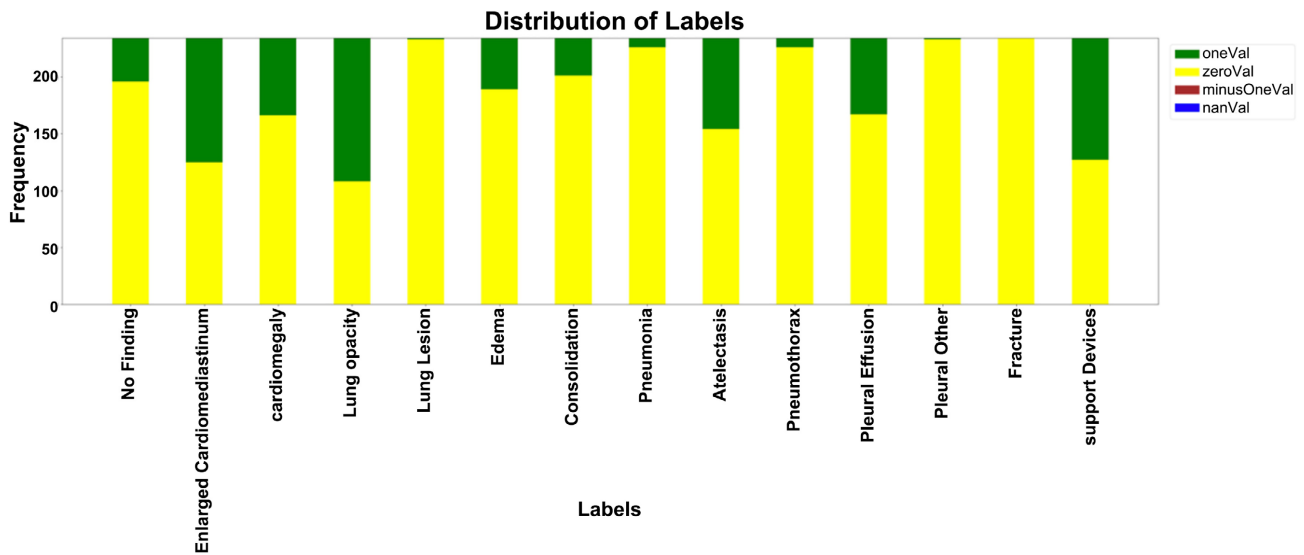

**Figure 8.** Distribution of test data and labels.

|    | Model Name | AUROC | Accuracy | F1 Score | Precision | Recall |
|----|-----------|-------|----------|----------|-----------|--------|
| 0  | CustomNet | 0.753760 | 0.863197 | 0.245288 | 0.389082 | 0.200749 |
| 1  | DenseNet121 | 0.783894 | 0.866112 | 0.278114 | 0.376906 | 0.240518 |
| 2  | ResNet50 | 0.730819 | 0.845245 | 0.183340 | 0.282960 | 0.169079 |
| 3  | Inception | 0.692110 | 0.829782 | 0.116774 | 0.165283 | 0.096235 |
| 4  | Vgg16 | 0.716072 | 0.838025 | 0.128222 | 0.186166 | 0.105025 |

**Figure 9.** Test metrics overall summary.

### 3.2. Test Metrics for Labels—AUROC

Densenet121 performed better in the case of individual labels also. Figure 10 shows AUROC values for various labels. Densenet121 model achieved AUROC ranging from 0.82 to 0.93 for the 5 labels those were part of the study conducted by Stanford ML group [1]. The performance was good for some other labels also like Pleural Other (AUROC: 0.97) and Lung Opacity (AUROC: 0.91). The worst performance was for Enlarged Cardiomediastinum (AUROC: 0.49).

### 3.3. Test Metrics for Labels—Accuracy

Densenet121 gave the best accuracy for individual labels. Figure 11 shows Accuracy values for various labels. Fracture, Lung Lesion, Pleural Other, Pneumonia and Pneumothorax achieved more than 95% accuracy. Enlarged Cardiomediastinum had the least accuracy of 53%.

| | | | | | AUROC |
| Model Name | CustomNet | DenseNet121 | Inception | ResNet50 | Vgg16 |
| Label | | | | | |
| --- | --- | --- | --- | --- | --- |
| Atelectasis | 0.784740 | 0.824351 | 0.750568 | 0.753490 | 0.778571 |
| Cardiomegaly | 0.757884 | 0.834515 | 0.745836 | 0.724309 | 0.707123 |
| Consolidation | 0.841248 | 0.897332 | 0.672396 | 0.826624 | 0.817126 |
| Edema | 0.882775 | 0.883598 | 0.811875 | 0.838566 | 0.788948 |
| Enlarged Cardiomediastinum | 0.538789 | 0.485358 | 0.512073 | 0.568607 | 0.520954 |
| Fracture | NaN | NaN | NaN | NaN | NaN |
| Lung Lesion | 0.047210 | 0.227468 | 0.085837 | 0.098712 | 0.287554 |
| Lung Opacity | 0.864051 | 0.911229 | 0.782775 | 0.823927 | 0.821355 |
| No Finding | 0.865602 | 0.850161 | 0.843582 | 0.854189 | 0.872180 |
| Pleural Effusion | 0.865135 | 0.925552 | 0.780409 | 0.833676 | 0.833318 |
| Pleural Other | 0.935622 | 0.969957 | 0.849785 | 0.884120 | 0.656652 |
| Pneumonia | 0.586836 | 0.682522 | 0.373341 | 0.605642 | 0.481195 |
| Pneumothorax | 0.632190 | 0.680310 | 0.681416 | 0.611173 | 0.830752 |
| Support Devices | 0.854588 | 0.882258 | 0.705276 | 0.748915 | 0.715652 |

Figure 10. Test labels comparison—AUROC.

| | | | | | Accuracy |
| Model Name | CustomNet | DenseNet121 | Inception | ResNet50 | Vgg16 |
| Label | | | | | |
| --- | --- | --- | --- | --- | --- |
| Atelectasis | 0.700855 | 0.752137 | 0.658120 | 0.675214 | 0.658120 |
| Cardiomegaly | 0.743590 | 0.713675 | 0.709402 | 0.709402 | 0.709402 |
| Consolidation | 0.858974 | 0.858974 | 0.858974 | 0.858974 | 0.858974 |
| Edema | 0.863248 | 0.863248 | 0.803419 | 0.811966 | 0.816239 |
| Enlarged Cardiomediastinum | 0.534188 | 0.534188 | 0.534188 | 0.534188 | 0.534188 |
| Fracture | 1.000000 | 1.000000 | 1.000000 | 1.000000 | 1.000000 |
| Lung Lesion | 0.995726 | 0.991453 | 0.995726 | 0.995726 | 0.995726 |
| Lung Opacity | 0.739316 | 0.782051 | 0.628205 | 0.756410 | 0.739316 |
| No Finding | 0.863248 | 0.807692 | 0.837607 | 0.858974 | 0.837607 |
| Pleural Effusion | 0.833333 | 0.871795 | 0.782051 | 0.773504 | 0.777778 |
| Pleural Other | 0.995726 | 0.995726 | 0.995726 | 0.995726 | 0.995726 |
| Pneumonia | 0.965812 | 0.965812 | 0.965812 | 0.965812 | 0.965812 |
| Pneumothorax | 0.965812 | 0.961538 | 0.965812 | 0.965812 | 0.965812 |
| Support Devices | 0.777778 | 0.769231 | 0.615385 | 0.679487 | 0.641026 |

Figure 11. Test labels comparison—Accuracy.

### 3.4. Confusion Matrix

Accuracy results may be misleading in some cases when the data set is unbalanced. We used confusion matrix to visualize true positives, false positives, true negatives, and false negatives as shown in Figure 12. Confusion Matrices for the best model, CustomNet and Densenet121 is given below.

### 3.5. CustomNet

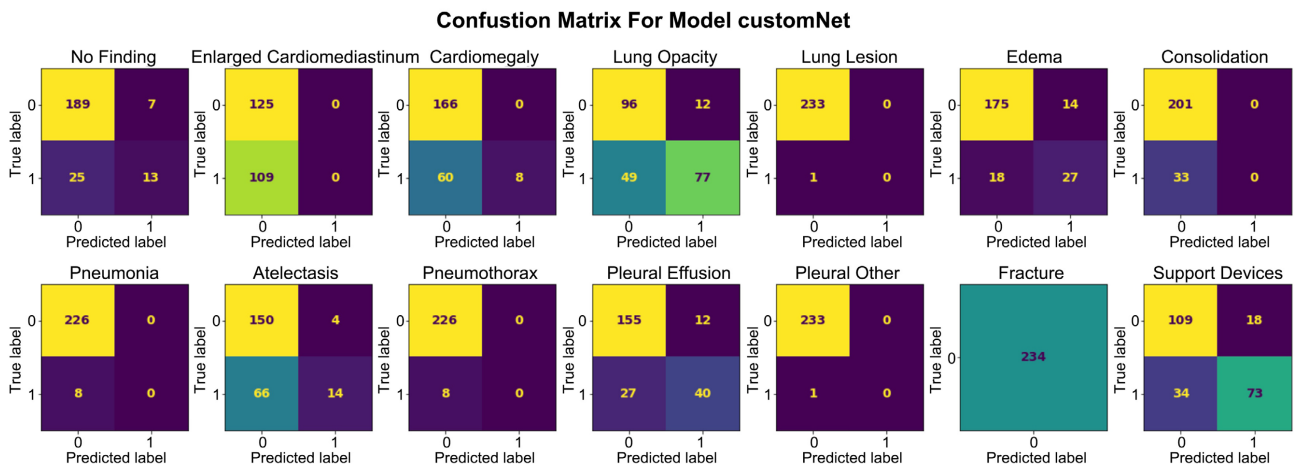

**Figure 12.** Confusion matrix for CustomNet.

### 3.6. DenseNet

Figure 13 demonstrates confusion matrix for DenseNet. The inability of the models to predict certain diseases are evident from the confusion matrix. For example, there were 68 positive cases for "Cardiomegaly" in the test data. Custom net predicted only 8 cases correctly where as DenseNet121 predicted 3 cases.

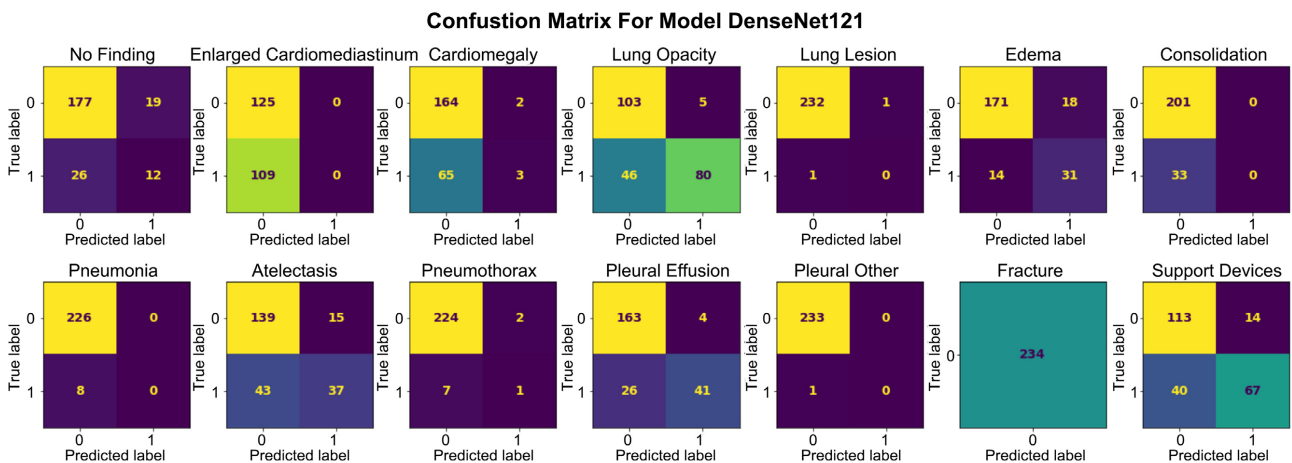

**Figure 13.** Confusion matrix for DenseNet.

## 4. Conclusion

The models were able to achieve ROC of about 0.78 and overall accuracy of about 87 percent. Dense121 pre-trained model gave the best performance for test data prediction. However, all models failed to accurately predict positive cases of certain diseases in spite of higher overall prediction accuracy rate. After examining the results, we hypothesize that these poor results are primarily due to lack of balanced training data. The number of positive cases in training data was very less compared to the negative class. Considering the uncertainty label as positive is not an effective approach for handling uncertainty in the dataset and is particularly ineffective on certain diseases. Overall, for 5 diseases, the best model was able to achieve AUROC comparable with baseline studies. Also model successfully predicted several other labels like Fracture, Lung Lesion, Pleural Other, Pneumonia and Pneumothorax with more than 95% accuracy.

### Optimization

We realize that the class imbalance affected model's ability to predict positive cases of certain labels. For example, Cardiomediastinum has only 4% positive cases in train data and Pneumonia has only 3% positive cases. In the future, we wish to address this by experimenting with over sampling or under sampling techniques. The bias towards the dominant class can be reduced by altering the training data in order to decrease the imbalance.

### Code Location

https://github.com/aravindsp/CS598_DL_Chest_X_RayClassification

### Notebook

https://github.com/aravindsp/CS598_DL_Chest_X_RayClassification/blob/main/code/cs-598-multi-labelchest-x-ray-classification.ipynb

### Python Script

https://github.com/aravindsp/CS598_DL_Chest_X_RayClassification/blob/main/code/cs-598-multi-labelchest-x-ray-classification.py

### Conflicts of Interest

The author declares no conflicts of interest regarding the publication of this paper.